\documentclass[aps,prl,twocolumn,showpacs]{revtex4}
\usepackage{amsbsy,latexsym}
\usepackage{amsmath}
\usepackage{amssymb, mathrsfs}
\usepackage[mathscr]{eucal}

\def\d{\operatorname{d}}\def\<{\langle}\def\>{\rangle}
\def\Tr{\operatorname{Tr}}\def\:{\hbox{\bf :}}
\def\spc#1{\mathcal{#1}}
\def\openone{1\!\!1}

\def\map#1{{\mathscr{#1}}}

\def\dim{\operatorname{dim}}

\def\qed{$\,\blacksquare$\par}

\def\dag{\dagger}
\def\geq{\geqslant}\def\leq{\leqslant}
 
\newtheorem{lemma}{Lemma}

\newtheorem{theorem}{Theorem}
\def\Proof{\medskip\par\noindent{\bf Proof. }}

\def\>{\rangle}
\def\<{\langle}

\begin{document}

\title{Quantum information becomes classical when distributed to many users} 
\author{Giulio Chiribella}\email{chiribella@fisicavolta.unipv.it} 
\affiliation{{\em QUIT Group}, Dipartimento di Fisica  ``A. Volta'' and INFM, via Bassi 6, 27100 Pavia, Italy}
\homepage{http://www.qubit.it}
\author{Giacomo Mauro D'Ariano}\email{dariano@unipv.it}
\affiliation{{\em QUIT Group}, Dipartimento di Fisica  ``A. Volta'' and INFM, via Bassi 6, 27100 Pavia, Italy}
\homepage{http://www.qubit.it}
\date{\today}

\begin{abstract}
  Any physical transformation that equally distributes quantum information over a large number $M$
  of users can be approximated by a classical broadcasting of measurement outcomes.  The accuracy of
  the approximation is at least of the order ${\mathcal O}(M^{-1})$. In particular, quantum cloning
  of pure and mixed states can be approximated via quantum state estimation. As an example, for
  optimal qubit cloning with 10 output copies, a single user has error probability $p_{err} \geq
  0.45$ in distinguishing classical from quantum output---a value close to the error probability of
  the random guess.
\end{abstract}
\pacs{03.67.Hk, 03.65.Ta}
\maketitle

Differently from classical information, which can be perfectly read out and copied, quantum
information cannot, since nonorthogonal quantum states can be neither perfectly distinguished
\cite{Helstrom76}, nor perfectly copied \cite{no-cloning}.  Since ideal distribution of quantum
information is impossible, one is then interested in the performance limits of optimal distribution,
and such interest has focused much attention in the literature to the problem of optimal cloning
\cite{OptCloning}. Optimal cloning consists in finding the physical transformation that converts $N$
copies of a pure state, randomly drawn from a given set, into the best possible approximation of $M
\geq N$ copies of the same state. More recently, the analogous problem for mixed states (optimal
broadcasting) has been considered \cite{Broad}.  In both cases of pure and mixed states, the optimal
transformation requires a coherent interaction of the input systems with a set of ancillae.  On the
other hand, classical incoherent schemes, such as the {\em measure-and-prepare}---where the $N$
initial copies are measured and $M$ copies of an estimated state are prepared---are suboptimal for
any finite $M$.

When cloning pure states, the measure-and-prepare scheme becomes
optimal in the asymptotic limit $M
\to \infty$ in all known kinds of cloning. This leads to conjecture that pure state cloning is
asymptotically equivalent to quantum state estimation \cite{ClonAndEst,Michael}, a conjecture
recently proved in Ref. \cite{Joonwoo}. Essentially, the line of proof is that a \emph{symmetric}
cloning transformation with $M = \infty$, when restricted to single clones, must be an entanglement
breaking channel, whence it can be realized by the measure-and-prepare scheme \cite{EBChannels}.
Such an argument, however, does not provide any estimate of the goodness of the classical scheme for
finite number $M$ of output copies, the situation of interest for applications and experiments.

In this letter we analyze the general class of quantum channels that equally distribute quantum
information to $M$ users, producing output states that are invariant under permutations. This class
contains cloning as a special case. We will show that for $M$ sufficiently large any channel of the
class can be efficiently approximated by a classical measure-and-prepare channel.  Indeed, we will
show that from the point of view of single users the states produced by the quantum and by the
classical channels are almost indistinguishable, with probability of error approaching the random
guess value $1/2$ at rate at least $\alpha/M$, $\alpha$ constant.  More generally, for any
group of $k$ users, the coherent and the incoherent schemes produce the same reduced state within an
accuracy $k\alpha/M$.  This also implies that entanglement between the output copies asymptotically
disappears at any given order $k$: for large $M$ only the $k$-partite entanglement with $k=\mathcal
O (M)$ can survive. The scaling $M^{-1}$ is a general upper bound holding for all physical
transformations that equally distribute quantum information among $M$ users, including pure state
cloning and mixed state broadcasting. Of course for specific
transformations the actual scaling can be even faster.

The mathematical description of a quantum channel that transforms states on the Hilbert space $\spc
H_{in}$ into states on the Hilbert space $\spc H_{out}$ is provided by a completely positive
trace-preserving map $\map E$. Since here we focus on channels that distribute quantum information
to $M$ users, we have $\spc H_{out} = \spc H^{\otimes M}$, with $\spc H$ denoting the single user's
Hilbert space. Moreover, since we require the information to be equally distributed among all users,
for any input state $\rho$ on $\spc H_{in}$ the state $\map E (\rho)$ must be invariant under
permutations of the $M$ output spaces.  Invariance under permutations implies that any group of $k$
users will receive the same state
\begin{equation}\label{RhoK}
\rho_{out}^{(k)}=
\Tr_{M-k}[\map E (\rho)]~,
\end{equation}  
$\Tr_{n}$ denoting partial trace over $n$ output spaces, no matter which ones.  In particular, each
single user receives the same state $\rho_{out}^{(1)}= \Tr_{M-1} [\map E (\rho)]$.  In the following,
we will name a channel with the above properties a \emph{channel for symmetric distribution of
  information} (\emph{SDI-channel}, for short).  Our goal will be to approximate any SDI-channel $\map E$
with a classical channel $\widetilde {\map E}$, corresponding to measure the input and
broadcast the measurement outcome, with each user preparing locally the same state accordingly.
Such channels have the special form
\begin{equation}\label{MeasAndPrep}
\widetilde {\map E} (\rho) = \sum_i~ \Tr[P_i \rho]~ \rho_i^{\otimes M}~,
\end{equation}      
where the operators $\{P_i\}$ represent the quantum measurement performed on the input ($P_i \geq 0,
\quad \sum_i P_i = \openone_{in}$), and $\rho_i$ is the state prepared conditionally to the outcome $i$.  The accuracy of the approximation is given by the trace-norm distance $|\!|\rho_{out}^{(1)} -\tilde \rho_{out}^{(1)}|\!|_1 =\Tr |\rho_{out}^{(1)} -\tilde
\rho_{out}^{(1)}|$ between the single user output states.  The trace-norm distance governs the
distinguishability of states \cite{Helstrom76}, namely the minimum
error probability $p_{err}$ in distinguishing between two equally
probable states $\rho_1$ and $\rho_2$ is given by
\begin{equation}\label{p_err}
p_{err}=\frac{1}{2}
-\frac{1}{4} |\!|\rho_1-\rho_2|\!|_1~,
\end{equation}
and for small distances it approaches the random guess value
$p_{err}=1/2$. In our case, a small distance $|\!|\rho_{out}^{(1)}
-\tilde \rho_{out}^{(1)}|\!|_1$ means that a single user has little
chance of distinguishing between the outputs of the two channels $\map
E$ and $\widetilde {\map E}$ by any measurement on his local state.
In addition, to discuss the multipartite entanglement in the state
$\rho_{out}^{(k)}$, we will consider the distance
$|\!|\rho_{out}^{(k)} -\tilde \rho_{out}^{(k)}|\!|_1$. Since the state
$\tilde \rho_{out}^{(k)}$ coming from $\widetilde {\map E}$ in
Eq.(\ref{MeasAndPrep}) is separable, a small distance means that any
group of $k$ users has a little chance of detecting entanglement.

The key idea of this letter is to get the approximation of SDI-channels exploiting the invariance of
their output states under permutations. In fact, permutationally invariant states have been thoroughly
studied in the research about quantum de Finetti theorem \cite{QDeFinetti}, where the goal is to
approximate any such state $\rho$ on $\spc H^{\otimes M}$ with a mixture of identically prepared
states $\tilde \rho = \sum_i~ p_i \rho_i^{\otimes M}$. In particular, as we will see in the
following, the recent techniques of Ref. \cite{OneAndHalf} provide a very useful tool to prove our
results. For simplicity, we will first start by considering the special case of SDI-channel with
output states in the totally symmetric subspace $\spc H_+^{\otimes M} \subset \spc H^{\otimes M}$,
which is the case, for example, of the optimal cloning of pure states.  Then, all results will be
extended to the general case of arbitrary SDI-channels.

In order to approximate channels we use the following finite version of quantum de Finetti theorem, which is proved with the same techniques of Ref.\cite{OneAndHalf}, with a slight improvement of the bound given therein \cite{Nota}:  
\begin{lemma} \label{lemma:SymmStates}
  For any state $\rho$ on $ \spc H_+^{\otimes M} \subset \spc H^{\otimes M}$, consider the separable
  state
\begin{equation}
\tilde \rho = \int \d \psi~  p(\psi)~ |\psi\>\<\psi|^{\otimes M}~,  
\end{equation}
where the probability distribution $p(\psi)$ is given by
\begin{equation}
p(\psi) =\Tr\left[ \Pi_{\psi} ~ \rho \right],\quad \Pi_{\psi}= d_M^+~|\psi\>\<\psi|^{\otimes M},
\end{equation}
where $\d \psi$ denotes the normalized Haar measure over the pure states $|\psi\> \in \spc H$, and
$d_M^+ =\dim (\spc H_+^{\otimes M})$. Then, one has
\begin{equation}\label{SqrtBound}
|\!| \rho^{(k)}  - \tilde \rho^{(k)}|\!|_1 \leq 4s_{M,k},\;\;\;
s_{M,k}\doteq 1- \sqrt{ \frac {d^+_{M-k}} {d_M^+}}~,
\end{equation}
$\rho^{(k)}$ denoting the reduced state $\rho^{(k)}= \Tr_{M-k}[\rho]$.
\end{lemma}    

\Proof  The identity in the totally symmetric subspace $\spc H^{\otimes n}_+ \subset \spc H^{\otimes n}$ can be written as
\begin{equation}\label{Res}
\openone^+_n= d^+_n \int \d \psi~ P_n(\psi)~,
\end{equation}
where $P_n(\psi) = |\psi\>\<\psi|^{\otimes n}$. Using Eq.\eqref{Res} with $n =M-k$, we  can write $\rho^{(k)} = d_+^{M-k}
\int \d \psi~ \rho_k (\psi)$, where
$\rho_k(\psi)= \Tr_{M-k} \left[\rho~ \openone^{\otimes k}
\otimes P_{M-k}(\psi) \right]$.  On the other hand, the reduced state
$\tilde \rho^{(k)}$ can be written as $\tilde \rho^{(k)} =
d_M^+ \int \d \psi~ P_k(\psi) ~ \rho_k(\psi) ~
P_k(\psi)$. Then, the difference between $\rho^{(k)}$ and $\tilde \rho^{(k)}$, denoted by $\Delta^{(k)}$, is given by
\begin{equation*}
 \Delta^{(k)} = d_{M-k}^+ \int \d \psi \left[ \rho_k (\psi) - \frac{d_M^+}{d_{M-k}^+}  P_k(\psi) \rho_k (\psi) P_k(\psi) \right].
\end{equation*}
Notice that the integrand on the r.h.s. has the form $A -BAB$, with $A(\psi)=\rho_{k}(\psi)$ and  $B(\psi)= \sqrt{d_M^+/ d_{M-k}^+}~ P_k(\psi)$.  Using the relation 
\begin{equation}
A -BAB= A (\openone -B) + (\openone -B) A -(\openone -B) A(\openone-B)
\end{equation}
we obtain 
\begin{equation}\label{Diff}
\Delta^{(k)} = d_{M-k}^+ \left( C  +  C^\dag -D\right)~,
\end{equation}
where 
\begin{eqnarray}
C&=& \int \d \psi~ A(\psi) \left[ \openone -B(\psi) \right]~,\\
\label{D} D &=& \int \d \psi~  \left[\openone -B(\psi) \right]~ A(\psi) ~\left[ \openone -B(\psi) \right]~.  
\end{eqnarray}
The operator $C$ is easily calculated using the relation       
\begin{equation*}
\begin{split}
\int \d \psi~ \rho_k (\psi)~  P_k(\psi) &=\int \d \psi \Tr_{M-k} [\rho ~ P_M(\psi)]\\  
&=\frac {\Tr_{M-k} [\rho ]} {d_M^+} = \frac  {\rho^{(k)}}{d_M^+}~,
\end{split}
\end{equation*}
which follows  from Eq. \eqref{Res} with $n=M$. In this way we obtain $C= s_{M,k}/d_{M-k}^+~\rho^{(k)}$~. Since $C$ is nonnegative, we have $|\!| C |\!|_1 = \Tr[C] =s_{M,k}/d_{M-k}^+$.
Moreover, due to definition \eqref{D} also $D$ is nonnegative, then we have $|\!| D |\!|_1 = \Tr
[D]= \Tr[C+ C^{\dag}]$, as follows by taking the trace on both sides of Eq.\eqref{Diff}. Thus, the norm of $D$ is $|\!|D|\!|_1 = 2|\!|C|\!|_1$. Finally, taking the norm on both sides of Eq. \eqref{Diff}, and using triangular inequality we get $|\!| \Delta^{(k)}|\!| \leq 4 d_{M-k}^+~|\!|C |\!|_1=4 s_{M,k}$, that is bound (\ref{SqrtBound}).  \qed

Since the dimension of the totally symmetric subspace $\spc
H_+^{\otimes n}$ is given by $d_n^+= \binom {d + n-1}{n}, ~d \doteq \dim (\spc H)$, for $M\gg k d$ the ratio
$d_{M-k}^+/d_M^+$ tends to $1- \frac{k(d-1)} M$.  Therefore, Lemma
\ref{lemma:SymmStates} yields
\begin{equation}\label{Bound0}
|\!| \rho^{(k)}  - \tilde \rho^{(k)}|\!|_1 \leq \frac{2(d-1)k}M ,\quad M\gg kd,
\end{equation} 
i.e. the distance between $\rho^{(k)}$ and the separable state $\tilde \rho^{(k)}$ vanishes as $k/M$. 

With the above lemma, we are ready to prove the approximation theorem for SDI-channels with output in the totally symmetric subspace:  
\begin{theorem}\label{theo:ApproxChannelBose}
Any SDI-channel $\map E$ with output states in the totally symmetric subspace $\spc H_+^{\otimes M} \subset \spc H^{\otimes M}$ can be approximated by a classical channel
\begin{equation}\label{ApproxChannel}
\widetilde{\map E} (\rho)= \int \d \psi~ \Tr[P_{\psi} \rho]~ |\psi\>\<\psi|^{\otimes M}~,
\end{equation}
where $P_{\psi}$ is a quantum measurement ($P_{\psi} \geq 0$ and  $\int \d \psi~P_{\psi}= \openone_{in}$). 
For large $M$, the accuracy of the approximation is 
\begin{equation}\label{Bound1}
|\!|\rho_{out}^{(k)} - \tilde \rho_{out}^{(k)} |\!|_1 \leq \frac{2 (d-1) k}M,\quad M\gg kd. 
\end{equation}
\end{theorem}
\Proof   Consider the channel ${\map E}^*$ in the Heisenberg picture, defined by the relation $\Tr[O \map E (\rho)]= \Tr[{\map E}^* (O) \rho]$ for any state $\rho$ on $\spc H_{in}$ and for any operator $O$ on $\spc H_{out}$. Since the channel $\map E$ is trace-preserving, ${\map E}^*$ is identity-preserving, namely ${\map E}^*(\openone_{out})= \openone_{in}$.   Applying Lemma 1 to the output state $\rho_{out}= \map E (\rho)$, we get $\tilde \rho_{out}= \int \d \psi~ \Tr[ \Pi_{\psi} \map E (\rho)]~|\psi\>\<\psi|^{\otimes M}$. Since $\Tr[ \Pi_{\psi} \map E (\rho)]= \Tr[{\map E}^* (\Pi_{\psi}) \rho]$, by defining  $P_{\psi} \doteq {\map E}^* (\Pi_{\psi})$, we immediately obtain that $\tilde \rho_{out}= \widetilde {\map E} (\rho)$, with $\widetilde{\map E}$ as in Eq. \eqref{ApproxChannel}. The operators $\{ P_{\psi}\}$ represent a quantum measurement on $\spc H_{in}$,  since  they are obtained by applying a completely positive identity-preserving map to $\Pi_{\psi}$, which is a measurement on $\spc H_{out}$. Finally,  the bound \eqref{Bound1} then follows from Eq. (\ref{Bound0}).   \qed

The above theorem proves that for large $M$ the quantum information
distributed to a single user can be efficiently replaced by the
classical information about the measurement outcome $\psi$.  In fact,
the single user output states of the channels $\map E$ and
$\widetilde{\map E}$ become closer and closer---and therefore less
distinguishable---as $M$ increases. For large $M$, the error
probability in distinguishing between $\rho_{out}^{(1)}$ and $\tilde
\rho_{out}^{(1)}$ has to satisfy the bound
\begin{equation}\label{p_errBound}
p_{err} \geq \frac 1 2 -\frac{ d-1} {2M}~,
\end{equation} 
namely it approaches $1/2$ at rate $M^{-1}$. For example, for qubits
Eq. (\ref{p_errBound}) gives already with $M=10$ an error probability
$p_{err}\geq 0.45$, quite close to the error probability of a purely
random guess.  More generally, the bound \eqref{Bound1} implies that
for any group of $k$ users there is almost no entanglement in the
state $\rho_{out}^{(k)}$, since it is close to a completely separable
state. 
As the number of users grows, multipartite entanglement vanishes at
any finite order: only $k$-partite entanglement with $k={\mathcal
O}(M)$ can survive.

Applying our approximation theorem to the particular case of pure
state cloning, we obtain a complete proof of its asymptotic
equivalence with state estimation. In fact, taking $\map E$ as an
optimal pure state cloning, the channel $\widetilde {\map E}$ yields
an approximation of $\map E$ based on state estimation (the
measurement outcomes of $P_{\psi}$ are in one to one correspondence
with the pure states on $\spc H$). On one hand, when applied to a pure
state $|\phi\>$, the optimal cloning gives fidelity $F_{clon}=\<\phi|
\rho_{out}^{(1)} |\phi\>$. On the other hand, since the measurement $P_{\psi}$
gives a possible estimation strategy, the fidelity of the
\emph{optimal} estimation $F_{est}$ cannot be smaller than $\<\phi|
\tilde \rho_{out}^{(1)} |\phi\>$. Therefore, the difference between
the two fidelities can be bounded as
\begin{equation}\label{Fclon}
0 \leq F_{clon} - F_{est} \leq |\!|\rho_{out}^{(1)} -\tilde \rho_{out}^{(1)}|\!|_1 \leq \frac{2
  (d-1)}{M},\quad M\gg d,      
\end{equation}
namely it approaches zero at rate $1/M$. A part from a constant, this is the optimal rate one can
obtain in a general fashion holding for any kind of pure state cloning. In fact, $1/M$ is the exact
rate in the case of universal cloning, where $F_{clon}-F_{est} = \frac{N(d-1)} {M(N+d)}$ (see
\cite{KeylClon} for the single-clone fidelity). In addition, from Eq. (\ref{Fclon}) it immediately
follows that any quantum cloning map for large numbers $N$ of input copies is approximated by state
estimation, since for cloning one has $M>N$, and $M$ is necessarily large. In this way we proved the
asymtotic equivalence beteween cloning and state estimation for any kind of cloning (see also the following
Theorem \ref{theo:ApproxChannel} for the general case of $\spc H_{out}\neq 
\spc H^{\otimes M}_+$), for either large $N$ or large $M$ (see open
problems in Ref. \cite{Michael}). We emphasize that the $M=\infty$ result of Ref. \cite{Joonwoo}
cannot be used to prove the large $N$ asymptotics.

All results obtained for SDI-channels with output in the totally symmetric subspace can be easily extended to arbitrary SDI-channels, exploiting the fact that any permutationally invariant state can be purified to a totally symmetric one \cite{OneAndHalf}:
\begin{lemma}\label{SymmetricPurification}
Any permutationally invariant state $\rho$ on $\spc H^{\otimes M}$ can be purified to a state
$|\Phi\> \in  \spc K^{\otimes M}_+\subset \spc K^{\otimes M}$, where $\spc K= \spc H^{\otimes 2}$.
\end{lemma}

Once the state $\rho$ has been purified, we can apply Lemma \ref{lemma:SymmStates} to the state $|\Phi\>$, thus approximating its reduced states. The reduced states of $\rho$  are then obtained by taking the partial trace over the ancillae used in the purification. 
This implies the following

\begin{lemma}\label{Lemma:DeFinettiEnt}
For any permutationally invariant state $\rho$ on $\spc H^{\otimes M}$, purified to $|\Phi\> \in  \spc K^{\otimes M}_+$, $\spc K = \spc H^{\otimes 2}$,  consider the separable state
\begin{equation}\label{SepState} 
\tilde \rho  = \int \d \Psi~  p(\Psi)~ \rho (\Psi)^{\otimes M}
\end{equation}
where  $\d \Psi$ is the normalized Haar measure over the pure states  $|\Psi\> \in \spc K$, $\rho( \Psi)$ is the reduced state $\rho(\Psi)= \Tr_{\spc H} \left[|\Psi\>\<\Psi|\right]$, and $p(\Psi)$ is the probability distribution given by $p(\Psi)=\Tr[\Pi_{\Psi} |\Phi\>\<\Phi|]$, with $\Pi_{\Psi}= D_M^+~|\Psi\>\<\Psi|^{\otimes M}$, $D_M^+ = \dim (\spc K_+^{\otimes M})$. Then, one has
\begin{equation}\label{Bound2}
|\!|\rho_{k,A} -\tilde \rho_{k,A} |\!|_1 \leq 4  S_{M,k},\quad S_{M,k} \doteq  1- \sqrt{\frac{ D_{M-k}^+}{D_M^+}}\ .
\end{equation}
\end{lemma}  
\Proof  Applying Lemma 1 to  $\tau= |\Phi\>\<\Phi|$, we get the state $\tilde \tau =  \int \d \Psi ~  p(\Psi)~|\Psi\>\<\Psi|^{\otimes M}$. The state $\tilde \rho$ is then obtained by tracing out the ancillae used in the purification, namely it is given by Eq.\eqref{SepState}.  Since  partial traces can only decrease the distance, the bound \eqref{Bound2} immediately follows from the bound \eqref{SqrtBound}. \qed        
It is then immediate to obtain the following:
\begin{theorem}\label{theo:ApproxChannel}
Any SDI-channel $\map E$ can be approximated by a classical channel
\begin{equation}
\widetilde{\map E} (\rho)= \int \d \Psi~ \Tr[P_{\Psi} \rho]~ \rho(\Psi)^{\otimes M}~,
\end{equation}
where $P_{\Psi}$ is a quantum measurement, namely $P_{\Psi} \geq 0$ and $\int \d \Psi~ P_{\Psi} =
\openone_{in}$.  
For large $M$, the accuracy of the approximation is
\begin{equation}\label{Bound2Bis}
|\!|\rho_{out}^{(k)} - \tilde \rho_{out}^{(k)} |\!|_1 \leq \frac{2(d^2-1) k}M,\quad M\gg kd^2. 
\end{equation}
\end{theorem}
This theorem extends Theorem \ref{theo:ApproxChannelBose} and all its consequences to the case of
arbitrary SDI-channels. In particular, it proves that asymptotically
the optimal cloning of mixed state can be efficiently simulated via
mixed states estimation. The results of the measurement
$P_{\Psi}$ are indeed in correspondence with pure states on $\spc H\otimes
\spc H$, and, therefore, with mixed states on $\spc H$. Accordingly,
the knowledge of the classical result $\Psi$ is enough to reproduce
efficiently the output of the optimal cloning machine. 

Notice the dependence on the dimension of the single user's Hilbert space in both Theorems
\ref{theo:ApproxChannelBose} and \ref{theo:ApproxChannel}: increasing $d$ makes the bounds
\eqref{Bound1} and \eqref{Bound2Bis} looser, leaving more room to cloning/broadcasting of genuine
quantum nature. Rather surprisingly, instead, the efficiency of our approximations \emph{does not
  depend} on the dimension of the full input Hilbert space, e.~.g. it doesn't depend on the number
$N$ of the input copies of a broadcasting channel. No matter how large is the physical 
system carrying the input information, if there are many users at the output there is no advantage of
quantum over classical information processing. Accordingly, our results can be applied to channels
from $\spc H^{\otimes N}$ to $\spc H^{\otimes M}$, even with $M <N$. As long as $M\gg kd^2$ any such
channel can be efficiently replaced by a classical one. In particular, this argument holds also for
the purification of quantum information \cite{cem,KeylPurif}: if $M$ is enough large, any strategy
for quantum purification can be approximated by a classical measure-and-prepare scheme. Only for
small $M$ one can have a really quantum purification.

In conclusion, we have considered the general class of quantum channels that equally distribute
information among $M$ users, showing that for large $M$ any such channel can be efficiently
approximated by a classical one, where the input system is measured and the measurement outcome is
broadcast, and each user prepares locally the same state accordingly.  The approximating channel can
be regarded as the concatenation of a \emph{quantum-to-classical} channel (the measurement),
followed by a \emph{classical-to-quantum} channel (the local preparation).  Actually, the latter
channel is needed only for the sake of comparison with the original quantum transformation to be
approximated, since, due to the data processing inequality, this additional stage can only decrease
the amount of information contained in the classical probability distribution of measurement
outcomes.  Therefore, asymptotically, there is no broadcasting of quantum information, but just an
announcement of the classical information extracted by a measurement. In synthesis, we cannot
distribute more information than what we are able to read out.


\par {\em Acknowledgments.---} This work has been founded by Ministero Italiano dell'Universit\`a e della Ricerca (MIUR)
through PRIN 2005.


\begin{thebibliography}{99}
\bibitem{Nota} The present bound on trace distance is asymptotically tighter than that of Ref.
  \cite{QDeFinetti}, improving  by a factor 2 the quality assessment of the channel approximation.
\end{thebibliography}
 \end{document}